\documentclass{article}
\usepackage{graphicx}

\title{A Reinforcement Learning System to Encourage Physical Activity in Diabetes Patients}

\author{
Irit Hochberg\\
Rambam Health Care Campus\\
Haifa, Israel\\
\texttt{i\_hochberg@rambam.health.gov.il}
\and
Guy Feraru\\
Faculty of Medicine\\
Technion, Israel Institute of Technology\\
Haifa, Israel\\
\texttt{guy.feraru@gmail.com}
\and
Mark Kozdoba\\
Faculty of Electrical Engineering\\
Technion, Israel Institute of Technology\\
Haifa, Israel\\
\texttt{mark.kozdoba@gmail.com}
\and
\and 
Shie Mannor \\
       Faculty of Electrical Engineering\\
       Technion, Israel Institute of Technology\\
       Haifa, Israel\\
       \texttt{shie@ee.technion.ac.il}
\and
Moshe Tennenholtz \\
       Faculty of Industrial Engineering and Management\\
       Technion, Israel Institute of Technology\\
       Haifa, Israel\\
       \texttt{moshet@ie.technion.ac.il}
\and 
Elad Yom-Tov\\
       Microsoft Research\\
       13 Shenkar st.\\
       Herzeliya, Israel\\
       \texttt{eladyt@microsoft.com}
}

\begin{document}

\maketitle

\begin{abstract}

Regular physical activity is known to be beneficial to people suffering from diabetes type 2. Nevertheless, most such people are sedentary. Smartphones create new possibilities for helping people to adhere to their physical activity goals, through continuous monitoring and communication, coupled with personalized feedback. 

We provided 27 sedentary diabetes type 2 patients with a smartphone-based pedometer and a personal plan for physical activity. Patients were sent SMS messages to encourage physical activity between once a day and once per week. Messages were personalized through a Reinforcement Learning (RL) algorithm which optimized messages to improve each participant's compliance with the activity regimen. The RL algorithm was compared to a static policy for sending messages and to weekly reminders. 

Our results show that participants who received messages generated by the RL algorithm increased the amount of activity and pace of walking, while the control group patients did not. Patients assigned to the RL algorithm group experienced a superior reduction in blood glucose levels (HbA1c) compared to control policies, and longer participation caused greater reductions in blood glucose levels. The learning algorithm improved gradually in predicting which messages would lead participants to exercise.

Our results suggest that a mobile phone application coupled with a learning algorithm can improve adherence to exercise in diabetic patients. As a learning algorithm is automated, and delivers personalized messages, it could be used in large populations of diabetic patients to improve health and glycemic control. Our results can be expanded to other areas where computer-led health coaching of humans may have a positive impact.

\let\thefootnote\relax\footnotetext{Summary of a part of this manuscript has been previously published as a letter in Diabetes Care, 2016.}

\end{abstract}

\section{Introduction}

Physical activity is highly recommended to patients with diabetes, since it is known that such activity leads to better control of glucose and other metabolic risk factors and improved quality of life \cite{colberg2010}. Despite recommendations, most diabetic patients fail to perform regular physical activity \cite{zhao2011}. A major objective of the caring medical team is to find better methods to encourage and incentivise physical activity in these patients. 

Apart from the obvious aim of improving persuasiveness in the communication between the patient and the healthcare providers on the issue of exercise \cite{kennedy2014}, there have been attempts to explore alternative approaches to improve adherence to physical activity in diabetic patients including financial incentives \cite{charness2009} and community programs \cite{pronk2015}. 

The smartphone revolution has brought entirely new opportunities for communicating with patients on a continuous basis and measuring movement, as well as other parameters, longitudinally. 

A very large proportion (30-70\%) of the population in developed and developing countries owns a smartphone \cite{economist2015}. In the last decade there have been multiple studies of mobile phone interventions using SMS messages to improve health related behaviors (reviewed in \cite{de2012}), and there are several previous studies that have tried to assess the effect of mobile phone applications in encouraging physical activity (reviewed in \cite{bort2014}). These studies use random messages or a display that quantitates the amount of activity performed.  None used a personalized learning algorithm to tailor messages to individuals. For example, two small-scale studies targeted patients with diabetes type 2 and took advantage of the ability of the patients' smartphone to recognize physical activity patterns \cite{aarsand2010,verwey2014}, but the feedback was either the count of number of steps walked, with no encouragement message, or a feedback provided by the nurse that cannot be scaled to a wide audience. 
The impact of wearable activity monitors (such as FitBit, Apple iWatch, and Microsoft Band) on encouraging exercise has not been assessed yet in an academic research setting.

The novel means of persistent communication afforded by smartphones, coupled with the ability to provide  reinforcement to patients as well as the almost immediate means to quantify its effect, has a potential to improve patient care on a wide scale. , but the use of personalized SMS messages that take into account the actual quantified behavior that needs to be reinforced has not been reported yet.  

Machine learning algorithms aim to discover a pattern, usually from previously-collected data. Reinforcement Learning (RL) algorithms \cite{gatti2015}, in contrast, are algorithms that learn by observing the result of an action taken by them and so can be applied in settings where data are scarce or varying. RL algorithms have been successfully applied in areas ranging from computer games \cite{tesauro1995} to health \cite{zhao2009}. In the latter, Paredes et al. \cite{paredes2014} used RL to select interventions to assist mildly depressed individuals, showing that RL-selected interventions were more effective than those selected using other strategies. Adaptive experimental design \cite{brown2009}, has been used to speed clinical trials and optimize treatment in a hospital setting \cite{shortreed2011}. Nevertheless, to the best of our knowledge, no previous work exists on the use of RL to help people adhere to medical recommendations in general, and to improve exercise in particular. 

The aim of this study is to assess the effectiveness of automatically-tailored, personalized feedback in increasing the adherence of diabetic patients to a personal physical activity regimen recommended by their diabetes specialist. We used a smartphone application that measured physical activity (especially walking) and sent feedback, in the form of SMS messages, to users. A learning algorithm, trained using the Reinforcement Learning framework, was used to predict the message most likely to increase activity on the following day. The primary outcome of this study was persistent improvement in physical activity. The secondary outcome was improved glycemic control. 

\section{Materials and Methods}

\subsection{Overview}

We developed a cellphone app which runs in the background of patients' smartphones and collects the amount of physical activity performed by patients. These data were transmitted to a central server. 

Each morning an RL algorithm assessed, for each patient, which SMS message would likely increase the physical activity of the patient in the upcoming day, and that message was sent to the patient. Users were represented to the RL algorithm by their demographics, past activity, expected activity, and message history.

The effectiveness of each message was assessed in the following morning, by calculating the amount of activity the patient performed since the last message was sent to him or her, and this signal served as the reward for training the RL algorithm. 

\subsection{Patient Characteristics}

Adult patients with type 2 diabetes were recruited for a 26-week-long study from the Endocrinology and Diabetes outpatient clinic at a tertiary hospital. Inclusion criteria were: non-optimal glycemic control (HbA1c\footnote{HbA1c is the common measure for control of blood glucose level in people with diabetes. It refers to the levels of glycated hemoglobin, a form of hemoglobin that is measured to identify the average plasma glucose concentration.} over 6.5\%); a sedentary lifestyle with no dedicated physical activity up to recruitment to the study; and ownership of an Android-based smartphone with a data connection (WiFi at home or cellular data). Exclusion criteria were other types of diabetes; any disability that precludes walking for 20 minutes. The study was approved by the Institutional Review Board of Rambam Health Care Campus. All patients gave written informed consent.

At recruitment, all participants received information on the importance of physical activity and a personal prescription for an activity plan including number of sessions of activity per week and time span for each per session (i.e., at least 2 hours of walking per week divided to 3 walking sessions per week). A dedicated smartphone app was installed on the participants' mobile phone. This application used the phone accelerometer to sense when participants were performing physical activity (defined as walking or running for 10 minutes or longer) and transmitted this information once every 2.5 hours to a central server. Feedback to patients was provided through SMS messages.  

To preserve battery life, the smartphone app sampled the accelerometer once every 3.5 minutes, and if walking was detected, kept monitoring the acceletometer until no walking was detected. Only contiguous walks of 10 minutes or more were collected, as shorter walks have a small effect on improvement in clinical outcomes.

Intensification of dietary or medical treatment was not restricted, when this was considered appropriate by the medical team. HbA1c measurements were performed by standard procedures before recruitment and every 3 months in the HMO lab of each subject. The patients filled a quality of life questionnaire \cite{jenkinson1993} before and after six months of participation in the study. They also filled a short questionnaire assessing satisfaction of the experience of using the application.

\subsection{Types of Feedback Messages}

Patients were randomized into a control arm and a personalized arm. The medical team was blinded to the type of messages each subject received. The control arm received identical unchanging once-weekly reminders to exercise. Patients in the personalized arm received daily feedback messages and weekly summaries. 

The daily feedback messages could be one of the following four messages:

\begin{enumerate}
	\item {\bf Negative feedback}: "You need to exercise to reach your activity goals. Please remember to exercise tomorrow".
	\item {\bf Positive feedback relative to self (Positive-Self)}: "You have so far achieved N\% of your weekly activity goal. Your exercise level is in accordance with your plan. Keep up the good work". 
	\item {\bf Positive feedback relative to others (Positive-Social)}: "You have so far achieved N\% of your weekly activity goal. You are exercising more than the average person in your group.  Keep up the good work".
	\item {\bf No message}.
\end{enumerate}

The percentage of the weekly goal ("N\%") was given as an integer greater than or equal to zero, computed according to the length of activity so far, compared to the length of activity expected given the exercise plan of the individual.

In general, messages did not necessarily reflect reality. For example, patients were not divided into groups, as is implies in the positive-social message.  Similarly, a negative message might be sent even though the patient has already achieved their activity goal. However, to allow the algorithm to learn a policy, we did not set constraints on the possible messages to be sent.

On most weeks the weekly summary message was "Please remember to exercise this week to reach your activity goals.". When patients achieved a significant exercise level, and not more than once per 3 weeks, they could receive one of the following messages:
\begin{enumerate}
	\item {\bf Maximal increase}: Over the past week you increased your activity more than at any previous week.
	\item {\bf Significant increase}: Over the past week you increased your activity more than most previous weeks.
	\item {\bf Maximal social}: You won the first place! Last week you increased your activity more than any other participant in the experiment.
	\item {\bf Significant social}: Last week you increased your activity more than most participants of the experiment.
\end{enumerate}

SMS messages were not sent to participants whose data were not received 12 hours or more prior to the current time, to reduce the chance that the system would send a message based on incorrect data.

\subsection{Feedback Message Policies}

After an initial period where feedback was sent according to a predetermined policy detailed below ("initial policy"), the decision on which daily feedback message to send was decided by a learning algorithm ("learned policy"). To allow the algorithm to collect information of outcomes to less likely feedback policies, exploration \cite{kaelbling1996} was used for messages that were deemed less likely to succeed such that they were sent with significant probabilities, as detailed below.

The initial algorithm (herewith referred to as "Initial policy") was set so that on 20\% of days, no message was sent. We then drew a uniform random number between 0 and 1. If that number was larger than the expected fraction of weekly activity at that day, the user would receive the negative feedback message. Otherwise, they would receive one of the positive messages, with equal probability. 

After a sufficient number of instances were collected, we implemented a learned decision mechanism for deciding on the feedback message. This mechanism received, for each user, the following set of attributes:

\begin{enumerate}
	\item Activity attributes:
				\begin{enumerate}
					\item Number of minutes of activity in the last day.
					\item Cumulative number of minutes of activity this week.
					\item Fraction of activity goal.
					\item Fraction versus expected at this point in the week.
				\end{enumerate}
	\item Demographics:
		\begin{enumerate}
			\item Age.
			\item Gender.
		\end{enumerate}
	\item Feedback attributes: Number of days since each feedback message was sent.
\end{enumerate}

Let $x_{i,t}$ denote a vector of the attributes above, for person $i$ at time $t$, and let $y_{i,t}$ denote the change in activity from day $t$ to day $t+1$ that is, the number of minutes of activity on day $t+1$ divided by those on day $t$. Following the Kesler construction \cite{duda1973}, we augment $x_{i,t}$ with an action vector $A$ such that the $i$-th element of $A$ is equal to 1 if and only if message $i$ was sent on day $t$.

The training data consists of all previously collected $x_{i,t}$ and $y_{i,t}$ pairs. We trained a learning algorithm, specifically a linear regression algorithm with interactions, to predict $y_{i,t}$ from $x{i,t}$. 

The learning algorithm was re-run every day and the most up-to-date model was used for prediction.

To predict the most appropriate action on day $\tau$, we apply the model to each $x_{i,\tau}$ and compute the resulting predicted $\hat{y_{i,\tau}}$. We then performed Bolzmann sampling \cite{duda2012} with $T_{Boltzmann} = 5$ on the outputs of the learning algorithm to choose the feedback message to be given. Thus, actions were chosen in proportion to their predicted effectiveness. This was done so that actions predicted not to be the best ones would still be tested, in addition to exploiting those actions predicted to be the best ones for the user.

There are many algorithms for addressing reinforcement learning problems. Most algorithms (Q-learning, TD learning, etc.) rely on either having access to the true underlying state, or to high quality features that represent the dynamics well. In our approach, we mostly tried to predict the effect of different actions on the immediate activity level given the current state of the patient rather than trying to change the patient's state. Thus, our policy is more of a "contextual bandit" type of algorithm \cite{li2010}. While we believe that introducing a state could be immensely useful, having statistical validity to it seems to require amounts of data beyond what we can expect.

\section{Results}

\subsection{Patient Characteristics}

A total of 27 patients were recruited, successfully installed the mobile app and transmitted data for at least one week. Patient characteristics are shown in Table \ref{tab:patients}. 

\begin{table}
\centering
\caption{Patient characteristics}
\label{tab:patients}
\begin{tabular}{|l|c|c|} \hline
Characteristic 	& Treatment & Control\\
\hline
Number							& 20				& 7 \\
Gender							& 8 female	& 1 females \\
Age	[yr]						& $58.7\pm2.1$& $55.1\pm3.6$ \\
Initial HbA1c	[\%]	& 7.7				& 8.7 \\
\hline\end{tabular}
\end{table}

\subsection{Application Use and Physical Activity Measured}

Target physical activity was on average 139 $\pm$ 62 minutes per week. The app continued to provide activity data for an average of 20.0 (SEM 1.6) weeks. Interrupted transmission was mostly due to change of the mobile phone or phone number during the study. Analysis was done on all participants that successfully initiated application use, including participants that did not complete the 26 weeks of the experiment. 

All patients reported that they did not perform regular physical activity prior to recruitment, but there is naturally no objective accelerometer data for the amount of activity performed before recruitment.  We decided that we could not separate the timing of providing the physical activity prescription from the recruitment process without causing any data collected in the first few weeks after recruitment to be biased.

\subsection{Effect of Different Messages Over Time}

We explored how each of the messages separately and how each two consecutive messages affected the change in the amount of activity, and found significant differences in the reaction of participants to different messages and message sequences. 

\begin{figure}[b]
\centering
\includegraphics[width=3in]{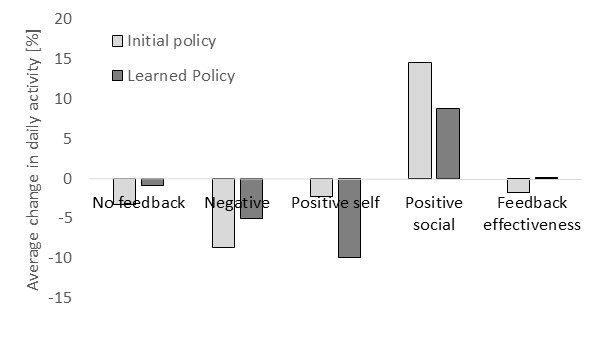}
\caption{Change in activity following feedback messages for the two feedback policies.}
\label{fig:daily}
\end{figure}

Figure \ref{fig:daily} shows the average improvement in activity ($y_{i,t}$) for each message, and the total change, weighted by the probability of each feedback message. The best increase in activity was found on the day after a positive social message, while negative messages and positive-self messages led to a decrease in the amount of activity. The differences in the change of activity between the initial policy and the learned policy were statistically significant (ANOVA, $P=0.0036$). 

\begin{figure}
\centering
\includegraphics[width=3in]{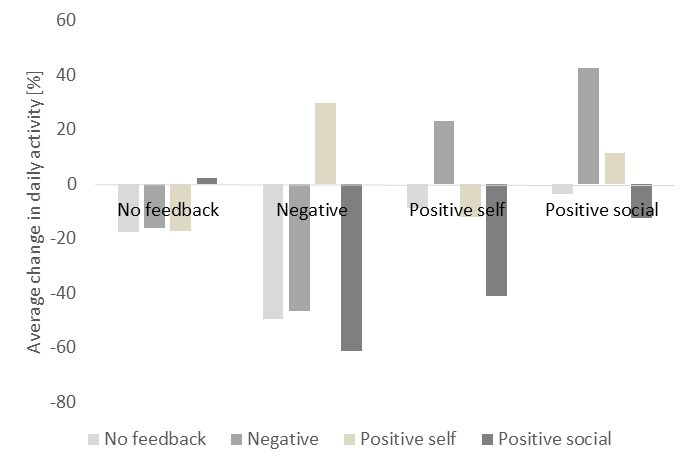}
\caption{Change in activity as a function of feedback, grouped by current feedback. Each group shows the average change in activity following the current feedback, given the previous feedback given to the user.}
\label{fig:pairs}
\end{figure}

One of the attributes given as inputs to the learned policy was the time since each feedback message was sent. This provides a limited form of historical context to the policy, allowing feedback to be dependent between days. Figure \ref{fig:pairs} shows the average improvement in activity for feedback on day N, given the feedback on the previous day (N-1). Differences in activity were statistically significant (ANOVA, $P=0.059$ for the previous action, and $P=0.021$ for the current action, $P=0.017$ for the interaction of the two actions). This figure demonstrates that such time-dependent feedback is more effective: For example, even though, on average, negative feedback produces a negative change in activity, it can, induce positive change if given before a positive-self feedback.  Similarly, positive social feedback is not effective if repeated day after day.

\subsection{Variability in patient response}

The average improvement in activity varies among patients. To demonstrate this, we represented each user according to the average change in their activity following each daily feedback message (i.e., a 4-dimensional vector). 

\begin{figure}[b]
\centering
\includegraphics[width=3in]{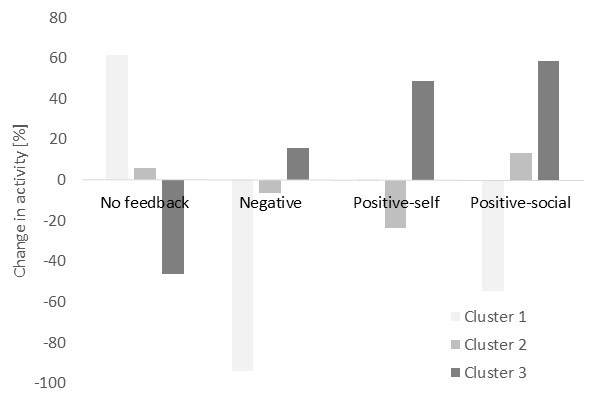}
\caption{Change in activity as a function of feedback message in each cluster. Cluster 1 comprised of 4 patients, cluster 2 of 9 patients, and cluster 3 of 5 patients.}
\label{fig:clustering}
\end{figure}

Figure \ref{fig:clustering} shows the results of clustering users, using k-means with 3 clusters. As the figure demonstrates, one group of patients (Cluster 1) reacted negatively to any feedback message. In contrast, patients in Cluster 3 reacted positively to messages, especially positive-social or positive-self. This demonstrates the importance of individually-tailored feedback delivered by our algorithm. 

Users in the different clusters differed in their demographics. Table \ref{tab:clusterdeomg} shows the percentage of females and the average age of patients in each cluster. As the Table shows, cluster 3, where patients reacted positively to messages (Figure \ref{fig:clustering}), is dominated by males. In contrast, cluster 2, where reactions to messages were overall weaker, consists of mostly women. Age variations are minor across clusters. Thus, there are significant correlates between patient gender and reaction to messages, demonstrating the importance of tailoring feedback according to these parameters, and to therefore providing them to the decision algorithm.

\begin{table}
\centering
\caption{Demographics of patients by cluster}
\label{tab:clusterdeomg}
\begin{tabular}{|l|c|c|c|} \hline
Demographic 	& Cluster 1 & Cluster 2 & Cluster 3\\
\hline
Percent female & 50 & 62 & 17 \\
Average age    & 57 & 54 & 56 \\
\hline\end{tabular}
\end{table}

\subsection{The Learning Process of the Algorithm Over Time}

\begin{figure}[h]
\centering
\includegraphics[width=3.0in]{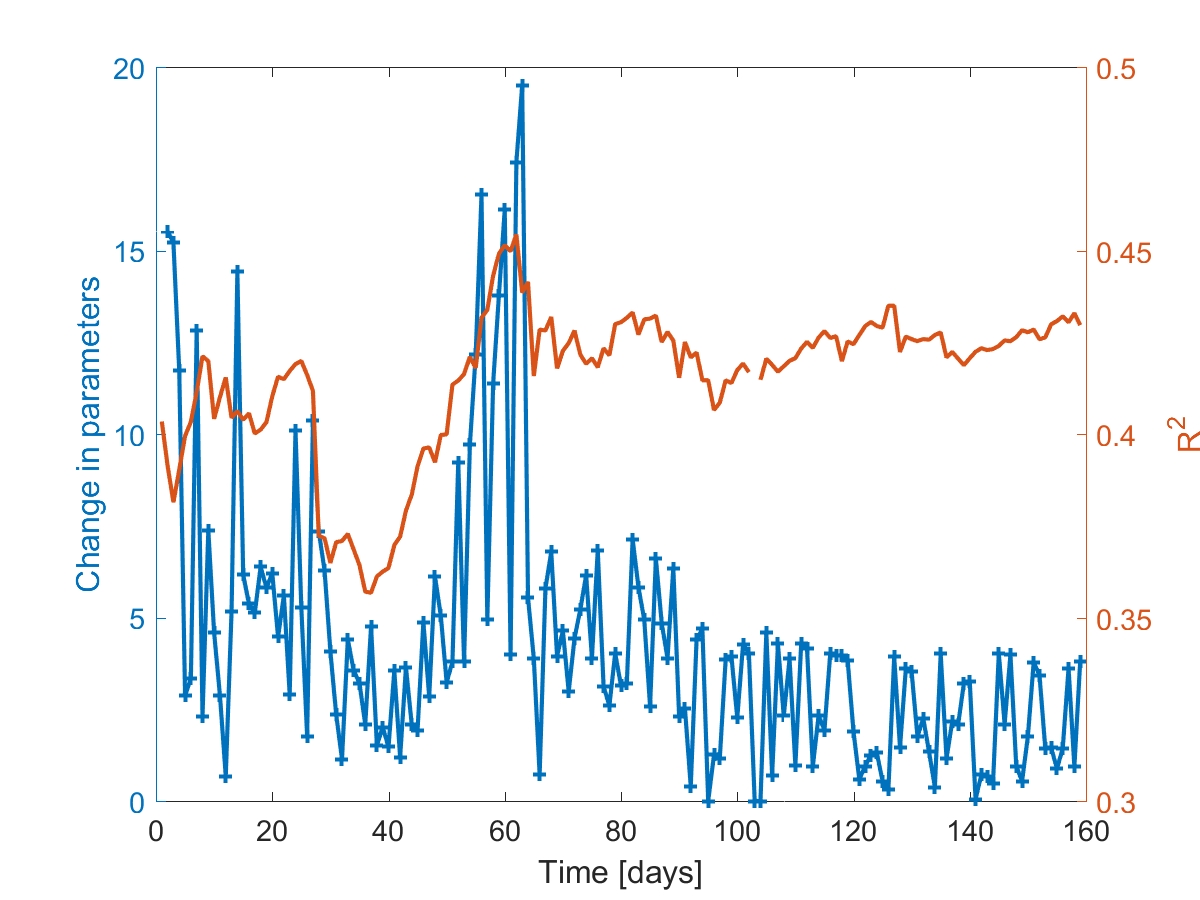}
\caption{Learning algorithm stability and predictiveness over time. The horizontal axis is time since the learning algorithm was applied to the experiment. The left vertical axis and the blue lines denoted by plus signs shows the change in algorithm parameters from day to day, and the right vertical axis and full brown line shows the $R^2$ value of the model.}
\label{fig:algoovertime}
\end{figure}

We investigated how the messages generated by the learning algorithm changed over time, as more information was collected on the response of the participants to feedback vis-a-vis their previous activity and demographics. Figure \ref{fig:algoovertime} shows how the learning algorithm gradually improves over time in predicting the amount of activity, demonstrating that much of the difference in exercise on a given day can be explained by the learning algorithm  indicating that much of  patient behavior is predictable.

Figure \ref{fig:algoovertime} shows the algorithm stability, calculated as the difference between the absolute values of the model parameters, and how much of the activity is explained by the predictions of the learning algorithm, as given by the adjusted $R^2$, over time.

First, we note that stability increases over time, as more data is collected. Second, $R^2$ initially increases, reaching approximately 0.43. This means that much of the difference in exercise on a given day can be explained by the learning algorithm attribute, indicating that to a large extent, patient behavior is predictable. We also note jumps in learning algorithm stability, for example around day 60. These jumps seem to correspond to major adverse weather events, and may be caused by new ways that people behave because of these events, creating unexpected data which cause the algorithm to learn a new hypothesis. This demonstrates the necessity to collect longitudinal data over wide-ranging circumstances, and possibly the need to include other  variables such as weather, calendar events, etc..

\subsection{Improvement in Activity Quantity and Walking Rate}

\begin{figure}
\centering
\includegraphics[width=3in]{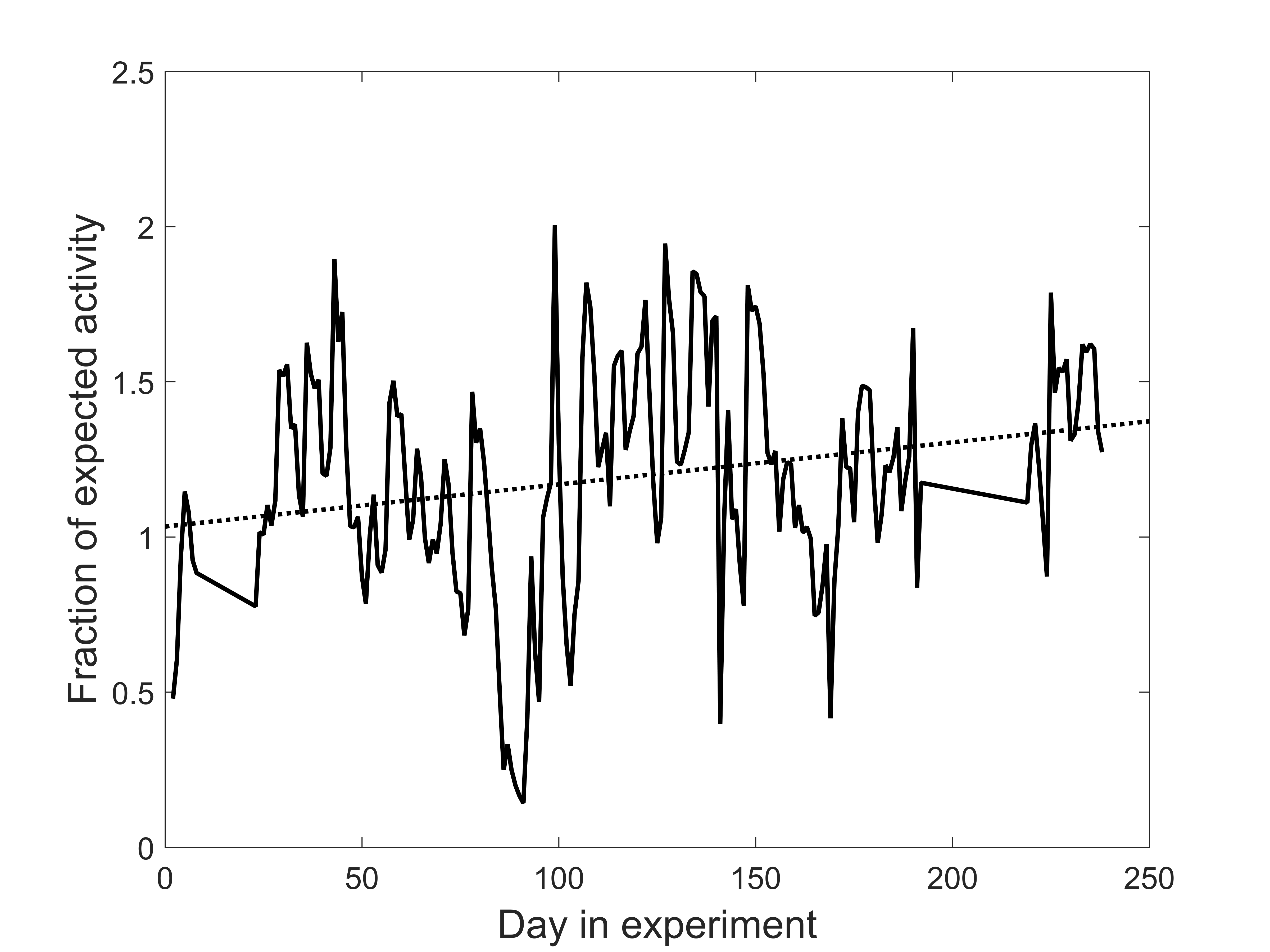}
\caption{The change in activity (shown as the fraction of the expected activity) over time for one sample user. The dotted line shows the linear slope of the curve.}
\label{fig:userovertime}
\end{figure}

We modeled the change in activity performed by patients over time (presented as fraction of target activity) using linear regression. Figure \ref{fig:userovertime} shows an example of the fraction of expected activity performed by one participant, together with the linear slope (which, for this patient, is equal to 0.0016) of this activity over the duration of the experiment. 

A linear function was fit for each participant separately, and the average slope for the participants in each policy group (weighted by the fit of the linear function) is shown in Table \ref{tab:slopes}. As the table shows, the slope of the learned policy was superior to both the control population and the initial policy. Whereas the latter two show a negative change in activity, the learned policy shows a positive slope, implying in increase in activity over time.

\begin{table}[t]
\centering
\caption{Rates of improvement in physical activity performed and in the rate of walking. In parenthesis, standard error of the mean. The slope of change in activity is measured by a linear fit to the plotted amount of daily exercise over time. The slope of the rate of walking is the change in the number of steps per minute during walking over time.}
\label{tab:slopes}
\begin{tabular}{|l|c|c|c|c|} \hline
Characteristic 	& \multicolumn{2}{|c|}{Treatment} & Control	\\
								& Initial 		& Learned 					& 				\\
\hline
Change in activity & -0.001		& +0.012						& -0.004  \\
$[min\;walking/day]$ & (0.008) & (0.002) & (0.002) \\
\hline
Change in rate of walking    & -0.009 & 0.002 & -0.010 \\
$[Hz/day]$										& (0.005) & (0.005) & (0.007) \\
\hline\end{tabular}
\end{table}

The rate of walking (steps per minute) was measured throughout the experiment. We modeled the change in the average weekly rate of walking over time using a linear model by fitting a linear function to the rate of walking for each participant separately over time, and the average slope for the participants in each policy group (weighted by the fit of the linear function) is shown in Table \ref{tab:slopes}. Patients in the control condition reduced their walking rate as the experiment progressed, consistent with the amount of walking they performed. In contrast, the personalized message population increased their walking rate over time significantly.

We analyzed the coefficients that affect the predictive ability of the learning algorithm and report those coefficients that had statistically significant values ($P<0.05$) in the linear model. These were:  

\begin{enumerate}
	\item The interactions between daily activity in the day before feedback is given and:	
		\begin{enumerate}
			\item The feedback message to provide.
			\item Activity performed so far.
			\item Time since each feedback message was given.
		\end{enumerate}
	\item The interactions between the activity performed so far and the time since each feedback message was given.
	\item The interactions between the fraction of activity performed so far and the time since each feedback message was given.
	\item The interactions between the time since each feedback message was given.
\end{enumerate}

\subsection{Change in Glycemic Control}

The initial HbA1c for all participant was 7.8 $\pm$ 1.0, and on average there was an improvement of 0.28 $\pm$ 0.84 in HbA1c for all patients. As mentioned, intensification of dietary or medical treatment was not restricted, so the change in HbA1c reflects a combination of the change in medical and dietary treatment and the change in exercise.

To assess the effect of variables of participation in the study on glycemic control, we constructed a linear model where the dependent variable is the difference between HbA1c levels at recruitment and the latest available measure of HbA1c. The independent variables are the number of days between measurements, initial HbA1c, and the activity target. Allocation to the personalized policy, higher initial HbA1c, and lower activity targets led to a superior reduction in HbA1c ($R^2=0.405$, $P<10^{-3}$).

\begin{figure}
\centering
\includegraphics[width=3in]{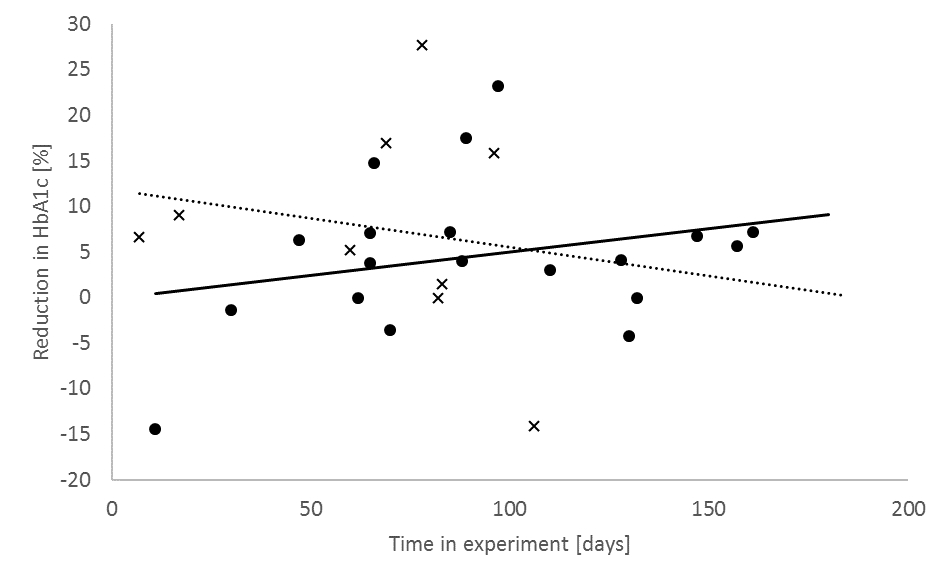}
\caption{Relative reduction in HbA1 over time. Dots represent measurements from people allocated to the personalized policy, and crosses the control policy. The dotted line is a linear fit to the control policy data, and the full line to the personalized policy.}
\label{fig:A1c_over_time}
\end{figure}

Let $HbA1c[t]$ be the blood glucose measure at time $t$. The relative reduction in HbA1c is given by: \[\frac{HbA1c[0] - HbA1c[t]}{HbA1c[0]}\] where the beginning of the experiment is at $t=0$. The relative reduction as a function of the time in the experiment can be seen in Figure \ref{fig:A1c_over_time}. The slope of a linear model for the treatment population is positive (0.05, $R^2=0.07$) while that of the control population is negative (-0.06, $R^2=0.03$), indicating that people in the treatment population experienced a reduction in blood glucose level the longer they participated and received messages determined by the personalized policy.

Thus, we conclude that receiving personal messages  is associated with a statistically significant reduction in HbA1c levels.

\subsection{Participant Satisfaction}

The results of the patient satisfaction questionnaire are shown in Table \ref{tab:survey}. Interestingly, both control and learned policy group participants reported increasing their physical activity. The learned policy population reported that the SMS messages helped them increase and maintain the level of their activity, significantly more than did the control population ($p<10^{-3}$). None of the participants in the control constant weekly reminder group felt that the SMS messages were helpful. Both groups said they received enough messages, though we interpret this result for the control population as unanimous lack of satisfaction with the unchanging wording of the feedback they received.

\begin{table}
\centering
\caption{Results of the patient satisfaction questionnaire. Only the response to the second question is statistically significantly different between control and personalized messages (chi2 test).}
\label{tab:survey}
\begin{scriptsize}
\begin{tabular}{|l|c|c|c|} \hline
Question 				& \multicolumn{2}{|c|}{Fraction answering "yes"} & P-value\\
								& Treatment & Control & \\
\hline
Did you increase your level of physical activity since joining the experiment? & 0.556 & 0.667 & 0.73 \\
Did the SMS messages help you increase the frequency of physical activity? & 0.800 & 0.000 & 0.01 \\
Did the SMS messages help you maintain your physical activity? & 0.875 & 0.333 & 0.07 \\
Do you think you received enough messages to improve your activity? & 0.778 & 1.000 & 0.46 \\
\hline\end{tabular}
\end{scriptsize}
\end{table}

\section{Discussion}

A large majority of patients with diabetes are resistant to the usual oral or written recommendations for physical activity they receive when encountering caregivers. Here we developed a system which takes advantage of the continuous monitoring and communication afforded by smartphones to explore an alternative approach for improving adherence. In this pilot study we evaluated the effect of feedback messages provided to patients directly by mobile phone based on their success in obtaining physical activity goals, as measured by a computerized mobile app. This requires careful integration of hardware, software, and human guidance. 

Our system used reinforcement learning to learn the feedback that will be most effective for each individual in any given situation, thus creating a highly personalized reminder service. Our results, as evident in the clusters of reactions to different feedback and the effect of message sequences, shows the importance of tailoring messages to each individual and at each time. 

We found that constant unvarying weekly reminders to perform physical activity are not effective in increasing activity, and that patients were not satisfied with receiving them. On the other hand, changing the messages based on the activity performed as determined by the learning algorithm was effective in increasing both the length of time walked and the rate of walking. Indeed, the RL algorithm learned to sequence messages to maximize efficiency. 

In our approach we learned a {\em single} model rather than a plurality of models. We ignored pertinent isseus such as the sex and age of the user. It stands to reason that building multiple models from data (e.g., one for women and one for men) could yield better results. Such an approach would require a larger population and would probably call for a different type of algorithm that takes into account contextual parameters as well leading to much better performance \cite{hallak2015}.

Our approach is fairly unique in that we do online traning within an experiment. In RL terminology, this is called on-policy learning. In  many treatments one must follow an off-policy scheme: collect data using one policy (usually a historical policy) and try to learn a new policy without actually executing it. This leads to several problems such as large variance and bias since exploration cannot be done where it matters most \cite{geist2014}. In our setting, this was not the case and we had the luxury of training and suusing the same policy.

Patients were satisfied with the experience of using the application when they received personalized messages generated by the algorithm. The length of participation and allocation to the learned policy group for which the learning algorithm was used were correlated with superior improvement in HbA1c over competing policies, namely, weekly reminders and policies which do not take into account the specific context and attributes of each user.

Our results suggest that this novel concept for increasing physical activity can be implemented economically, efficiently and effectively, leading to desired highly positive results. Notice that our approach not only allows for a predictive tool (going beyond current messaging systems), but also provides a method for personalized care.

This small-scale study demonstrates the general concept that continuous monitoring and personalized guidance generated by a computer can have a significant impact on patient behavior. Unlike many current e-medicine systems that require input from the patient or the healthcare provider, the use of an automatic algorithm can be applied to very large groups of subjects. We plan to expand our result to the even more general concept that digitally-generated health coaching of humans can have a positive impact. Further studies in larger scale and longer periods of time are needed to evaluate whether the digital revolution and the potential to directly communicate with large groups of subjects and assess the actual behavior reinforced can lead to a major improvement in their health-related behaviors or in their actual health.

\bibliographystyle{abbrv}
\bibliography{RLDiabetes}  

\end{document}